\documentclass[square]{ws-procs975x65}
\pdfoutput=1
\usepackage{graphicx}
\usepackage{color}
\usepackage{epsfig}
\usepackage{dcolumn}
\usepackage{subfigure}
\usepackage{amsmath,amssymb}
\usepackage{array}

\newcommand{\TeV}{\text{\,TeV}}
\newcommand{\GeV}{\text{\,GeV}}

\newcommand{\be}{\begin{equation}}
\newcommand{\ee}{\end{equation}}
\newcommand{\beq}{\begin{eqnarray}}
\newcommand{\eeq}{\end{eqnarray}}
\newcommand{\bpm}{\begin{pmatrix}}
\newcommand{\epm}{\end{pmatrix}}

\begin{document}

\title{Top-seesaw assisted technicolor model with $126$ GeV Higgs boson\footnote{This talk is based on \cite{Fukano:2012nx} and given at 2012 Nagoya Global COE workshop `` Strong Coupling Gauge Theories in the LHC Perspective'' (SCGT 12), December 4-7, 2012}}
\author{Hidenori S. Fukano\footnote{speaker,E-mail : hidenori.f.sakuma@jyu.fi} and Kimmo Tuominen}
\address{Department of Physics, University of Jyv\"askyl\"a, P.O.Box 35, FIN-40014 Jyv\"askyl\"a, Finland \\
Helsinki Institute of Physics, P.O.Box 64, FIN-00014 University of Helsinki, Finland\\}

\begin{abstract}
We discuss a model which involves the top quark condensation and the walking technicolor. We focus on the scalar boson in such a model from the viewpoint of the observed scalar boson at the LHC.
\end{abstract}


%
\section{Introduction}
%
The ATLAS and CMS collaborations at the LHC have announced the discovery of a new scalar boson with mass $m_h\simeq 126$ GeV \cite{:2012gk,:2012gu}. The production cross section and decay rates of this new boson appear to be consistent with the prediction of the Higgs boson in the Standard Model (SM). Therefore the next logical step is to try to uncover its properties more precisely and to see how well it fits in with various extensions of the SM. A scenario based on the dynamical electroweak symmetry breaking still remains as a viable alternative, although the discovery of a light scalar boson is a severe obstruction for traditional technicolor models \cite{Susskind:1978ms}. 

There are generally at least two different alternatives based on the dynamical electroweak symmetry breaking. First possibility is walking technicolor which is controlled by the quasiconformal gauge dynamics \cite{Holdom:1984sk}. The walking technicolor model has an approximate scale symmetry and a scalar boson emerges as the pseudo Nmabu-Goldstone boson after the approximate scale symmetry is broken. This scalar boson, so-called techni-dilaton, might be light and 
explain the observed new boson at the LHC \cite{Matsuzaki:2012mk,Elander:2012fk}. Second possibility is the top quark condensation \cite{Miransky:1988xi,Bardeen:1989ds} where the electroweak symmetry breaking is triggered by the Nambu-Jona-Lasinio (NJL) dynamics \cite{Nambu:1961tp}. The NJL dynamics leads to emergence of 
a scalar boson with mass $m_h = 2 \Sigma$ where $\Sigma$ is the dynamical fermion mass.  
This scalar boson is described as a SM like electroweak scalar doublet. This is different from the description of the techni-dilaton  \cite{Matsuzaki:2012mk}.

In this talk we consider the new observed scalar boson from the viewpoint of the top quark condensation model. In the top quark condensation model, scalar boson is composed of the top quark and its mass is related to the top quark mass by the NJL dynamics, i.e. $m_h = 2 m_t$. This is not suitable to explain the observed scalar boson with $m_h \simeq 126 \GeV$ at the LHC. However, it might be possible to explain the observed scalar boson mass in the top quark condensation model by sharing the top quark mass with another dynamical sector, e.g. extended technicolor \cite{Dimopoulos:1979es}. This scenario is the top-seesaw assisted technicolor model \cite{Fukano:2012nx,Fukano:2011fp} where we have used the top-seesaw model \cite{Dobrescu:1997nm} which might be a promising model among several models based on the top quark condensation under the present experimental constraints.

%
\section{Top-seesaw assisted technicolor (TSSTC) model}

In this section, we consider the $126 \GeV$ scalar boson in a model which involves walking technicolor and top-seesaw dynamics simultaneously. However, in this talk, we concentrate only on the top-seesaw sector to focus on a scalar boson originating from the electroweak doublet like the Higgs boson in the SM. Particle contents are summarized in table.\ref{TSScontents}. 
Here $Q^{1,2,3}$ are the usual SM chiral quarks, but $U^{(4)},D^{(4)}$ are vector-like, i.e. the electroweak singlets. 
The usual SM leptons are sufficient to avoid the gauge anomalies, and we do not show them explicitly in the table.\ref{TSScontents}. 
\begin{table}
\begin{center}
\tbl{Particle content and charge assignments.\label{TSScontents}}
{
\begin{tabular}{| c | c | c | c |c|}
\hline
field  & SU(3)$_1$  & SU(3)$_2$ & SU(2)$_L$ &$U(1)_Y$\\ 
\hline 
\, & \, & \, & \, & \, 
\\[-2.5ex]
\hline
$Q^{(3)}_L$ &3 & 1 & 2 & 1/6 
\\
$U^{(3)}_R , D^{(3)}_R$ &1 & 3 & 1 & (2/3 , -1/3) 
\\
\hline 
\, & \, & \, & \, & \, 
\\[-2.5ex] 
\hline
$U^{(4)}_L , D^{(4)}_L$ &1 & 3 & 1 &(2/3, -1/3)  
\\
$U^{(4)}_R , D^{(4)}_R$ &3 & 1 & 1 & (2/3,-1/3) 
\\
\hline 
\, & \, & \, & \, & \, 
\\[-2.5ex] 
\hline 
$Q^{(1,2)}$ &1 & 3 & SM & SM
\\
\hline
\end{tabular}
}
\end{center}
\end{table}%
We assume that $SU(3)_1$ topcolor gauge coupling is stronger than $SU(3)_2$ topcolor gauge coupling and the topcolor breaking, $SU(3)_1 \times SU(3)_2 \to SU(3)_c$, is triggered at a scale $\Lambda$. The unbroken  $SU(3)_c$ is the usual color gauge group. The topcolor breaking provides following four fermion interactions at scale $\Lambda$:
\beq
{\cal L}^{4f}
=
G_b \left( \bar{D}^{(4)}_R Q^{(3)}_L\right)^2 
+
G_t \left( \bar{U}^{(4)}_R Q^{(3)}_L\right)^2 
+
G_{tb} \left( \bar{Q}^{(3)}_L U^{(4)}_R \right) 
\left( \bar{D}^{(4) c}_R i \tau_2 Q^{(3) c}_L\right)
\,,
\label{4f-present}
\eeq
where the superscript $^c$ implies charge conjugation. The diagonal terms, $G_b$ and $G_t$ arise from the exchange of color octet massive gauge bosons with mass $\sim\Lambda$ originating from the topcolor breaking. The off diagonal term $G_{tb}$ may arise from e.g. the topcolor instantons \cite{He:2001fz}. Using the fermion bubble sum approximation \cite{Bardeen:1989ds}, the low energy Lagrangian at $\mu < \Lambda$ is given by
\beq
{\cal L}_{\rm TSS} (\Phi_1,\Phi_2)
=
\sum_{i=1,2} \left| D_\mu \Phi_i \right|^2 
+ {\cal L}_M
+ {\cal L}^{\rm TSS}_{\rm yukawa}
- V_{\rm TSS}(\Phi_1,\Phi_2) \,,
\label{TSS-EFT}
\eeq
where ${\cal L}^{\rm TSS}_{\rm yukawa}$ consists of the Yukawa interaction terms for the third family quarks and their vector-like partner quarks and is given explicitly by
\beq
{\cal L}^{\rm TSS}_{\rm yukawa}
=
-  y_1 \bar{Q}^{(3)}_L \Phi_1 D^{(4)}_R 
- y_2 \bar{Q}^{(3)}_L \tilde{\Phi}_2 U^{(4)}_R 
+ {\rm h.c. }
\,.
\label{reno-yukawa-TSS}
\eeq
The potential $V_{\rm TSS}(\Phi_1,\Phi_2) $ is given by
\beq
V_{\rm TSS}(\Phi_1,\Phi_2) 
&=& 
M^2_{11} |\Phi_1|^2 + M^2_{22} |\Phi_2|^2 
- M^2_{12} \left[ \Phi^\dagger_1 \Phi_2 + {\rm h.c.}\right] 
\\[1ex]
&&
+ \frac{1}{2} \lambda_1( \Phi^\dagger_1 \Phi_1 )^2 
+ \frac{1}{2} \lambda_2 ( \Phi^\dagger_2 \Phi_2)^2 
+ \lambda_3(\Phi^\dagger_1 \Phi_1)(\Phi^\dagger_2 \Phi_2)
+ \lambda_4(\Phi^\dagger_1 \Phi_2)(\Phi^\dagger_2 \Phi_1)
\,.\nonumber
\label{TSS-higgs-potential}
\eeq
Finally, ${\cal L}_M$ in Eq.(\ref{TSS-EFT}) is the electroweak singlet mass term and is given by
\beq
{\cal{L}}_M=
-\bar{U}^{(3)}_R M^U_{34} U^{(4)}_L - \bar{U}^{(4)}_R M^U_{44} U^{(4)}_L 
+\left[ U \to D \right] + \text{h.c.}\,.
\eeq
The doublets $\Phi_{1,2}$ are parametrized as $\Phi_i = \bpm \pi^+_i \,,\,\left( v_i + h^0_i - i \pi^0_i \right)/\sqrt{2}\epm^T$ $(i=1,2)$ and the covariant derivatives for $\Phi_i$ are of the same form as for the SM Higgs doublet. %
The NJL dynamics based on Eq.(\ref{4f-present}) is rewritten by the renormalization group equations for $y_i,\lambda_i$ in Eq.(\ref{reno-yukawa-TSS}) together with suitable compositeness conditions \cite{Bardeen:1989ds,He:2001fz}. %
\begin{figure}[htbp]
\begin{center}
\begin{tabular}{cc}
{
\begin{minipage}[t]{0.5\textwidth}
\includegraphics[scale=0.67]{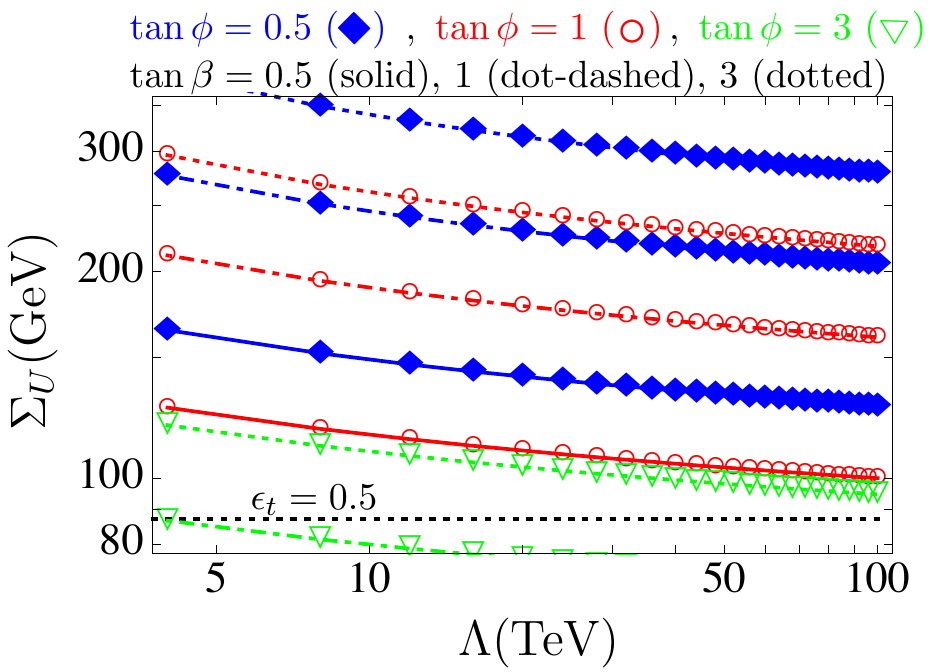} 
\end{minipage}
}
{
\begin{minipage}[t]{0.5\textwidth}
\includegraphics[scale=0.7]{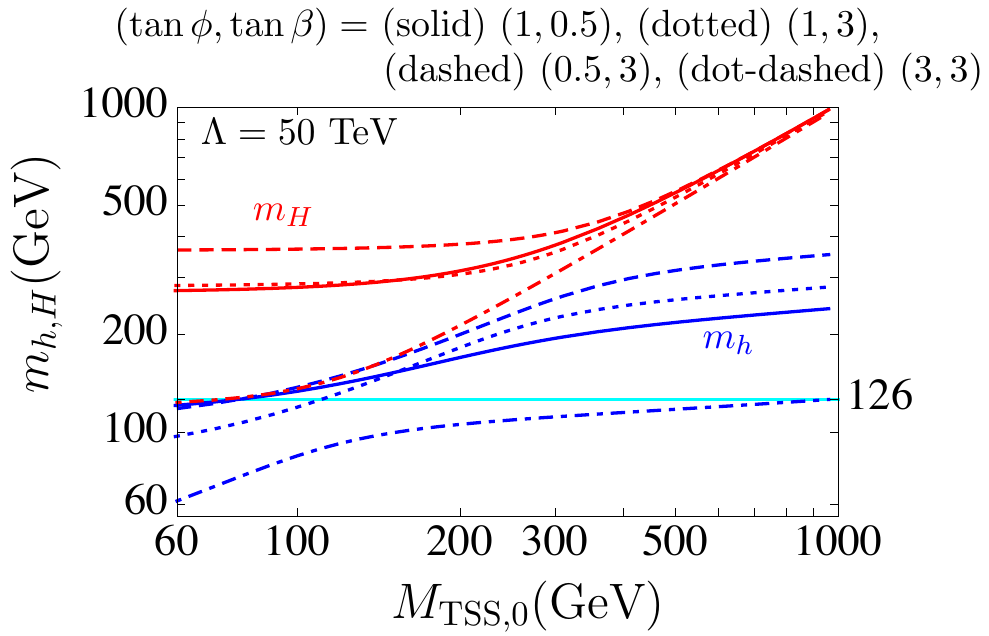} 
\end{minipage}
}
\end{tabular}
\caption[]{
The dynamical mass for (left) the top quark sector and (right) the CP-even higgs boson with $\tan \phi  = 0.5,1,3$ and $\tan \beta = 0.5,1,3$ . The horizontal dotted line in the left panel corresponds to  $m_t(\text{TSS})=0.5 m_t$. The horizontal line in the right panel corresponds to $m_h = 126 \GeV$.\label{fermion-dynamical-mass}}
\end{center}
\end{figure}%
In Fig.\ref{fermion-dynamical-mass} the resultant dynamical (left) fermion mass and (right) CP-even Higgs bosons are shown, and $\tan \beta\,,\tan \phi$ are given by $\tan \beta \equiv v_2/v_1$, $\tan \phi \equiv v_{\rm TC}/\sqrt{v^2_1 + v^2_2}$ and $v^2_{\rm EW} = (246 \GeV)^2 \equiv v^2_1 + v^2_2 +v^2_{\rm TC}$. The dynamical fermion mass for the top quark sector is $\Sigma_U\equiv y_2 v_2/\sqrt{2}$. The parameter $M_{{\rm TSS},0}$ is defined as $M_{{\rm TSS},0} \equiv M^2_{12}/(\sin \beta \cos \beta)$ and $\epsilon_t \equiv m_t({\rm ETC})/m_t$ where $m_t({\rm ETC})$ is contribution to the top quark mass from the extended technicolor sector. We fix $\epsilon_t = 0.5$ here. %
From the right panel in Fig.\ref{fermion-dynamical-mass}, we find several candidates of  parameters to realize the $m_h = 126 \GeV$, for example,
\beq
M_{{\rm TSS},0} = 77 \GeV && \text{for} \quad \tan \phi = 1 \quad , \quad \tan \beta =0.5 
\,, \label{MTSS0-sample0}
\\
M_{{\rm TSS},0} = 960 \GeV \!\!&& \text{for} \quad \tan \phi = 3 \quad , \quad \tan \beta =3 \,. \label{MTSS0-sample3}
\eeq
The case of Eq.(\ref{MTSS0-sample0}) leads to a problematic mass difference among physical Higgs bosons: $m_h,m_A \simeq 126 \GeV < m_H \simeq m_{H^\pm} \simeq 300 \GeV$ where $A,H^\pm$ are the CP-odd and charged Higgs boson respectively. This mass difference  gives a large contribution to the Peskin-Takeuchi $T$-parameter \cite{Peskin:1990zt} and it is disfavored by the electroweak precision tests. On the other hand, the case Eq.(\ref{MTSS0-sample3}) provides $m_h \simeq 126 \GeV < m_H \simeq m_A \simeq m_{H^\pm} \simeq 1 \TeV$ which is not constrained by the electroweak precision tests, since the degenerate heavy higgs bosons do not result in a large contribution to the $T$-parameter. Furthermore $R_b \equiv \Gamma(Z \to b\bar{b})/\Gamma(Z \to \text{hadrons})$ constraint also favors the case of Eq.(\ref{MTSS0-sample3}) rather than 
Eq.(\ref{MTSS0-sample0}) \cite{Fukano:2012nx}. %
Therefore, we conclude that the case of Eq.(\ref{MTSS0-sample3}) is favorable parameter set in the present model from the viewpoint of both the electroweak precision tests and the observed new scalar boson mass.

%
\section{$126 \GeV$ Higgs boson in the TSSTC at the LHC}
%

\begin{figure}[htbp]
\begin{center}
\begin{tabular}{cc}
{
\begin{minipage}[t]{0.5\textwidth}
\includegraphics[scale=0.6]{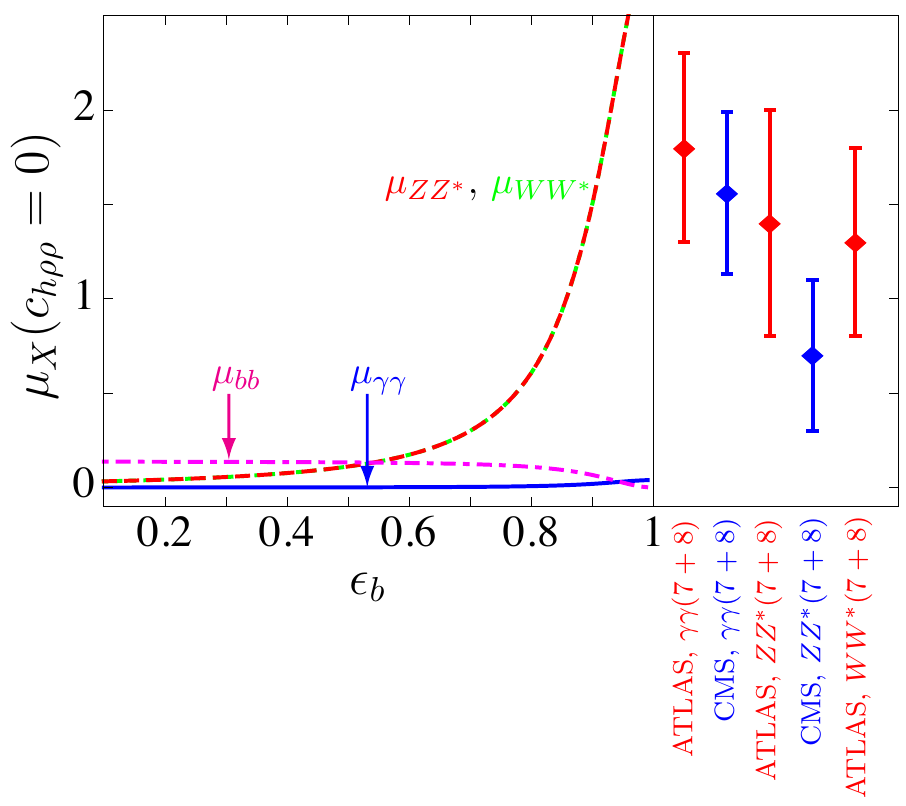} 
\end{minipage}
}
{
\begin{minipage}[t]{0.5\textwidth}
\includegraphics[scale=0.6]{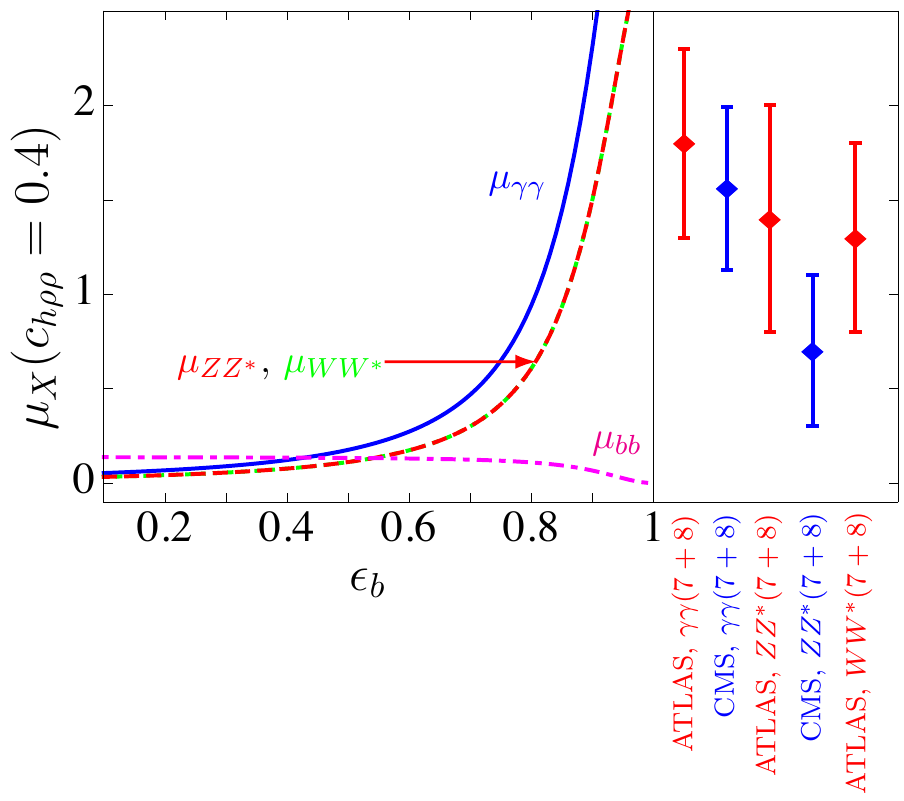} 
\end{minipage}
}
\end{tabular}
\caption[]{
The signal strength $\mu_X (X = \gamma\gamma,WW^*,ZZ^*,bb)$ as a function of $\epsilon_b$ in the present model. The blue solid, green dotted, red dashed, magenta dot-dashed curves correspond to $\mu_{\gamma \gamma}, \mu_{WW^*},\mu_{ZZ^*}, \mu_{bb}$, respectively. In both panels, the LHC combined results for $\gamma \gamma\,,\,ZZ^*\,,\,WW^*$ in\cite{:2012gk,:2012gu,CMS-PAS-HIG-12-015,CMS-PAS-HIG-12-016} are shown together. The left panel shows the signal strength for a case with $C_{h\rho\rho} = 0$ and the right panel shows the signal strength for a case with $C_{h\rho\rho} = 0.4$ with $M_\rho = 1 \TeV$ 
\label{higgs-LHC}}
\end{center}
\end{figure}

In this section, we compare the Higgs boson in the present model for Eq.(\ref{MTSS0-sample3}) to the recent LHC SM Higgs boson search data. For this purpose, we vary $\epsilon_b \equiv m_b({\rm ETC})/m_b$ in a range $0.1 \leq \epsilon_b \leq 1$ which does not affect the previous discussions. %
We focus on the signal strength defined as
\beq
\mu^{\rm ggF}_X \equiv \frac{\sigma_{\rm ggF}(h_{\rm TSSTC})}{\sigma_{\rm ggF}(h_{\rm SM})} \times \frac{{\rm Br}(h_{\rm TSSTC} \to X)}{{\rm Br}(h_{\rm SM} \to X)}
\,,\label{signalstrength-ggF}
\eeq
where $\sigma_{\rm ggF}$ is the production cross section of the Higgs boson by the gluon fusion and ${\rm Br}(h \to X)$ $(X=\gamma\gamma,WW^*,ZZ^*)$ is the branching ratio of the Higgs boson. Furthermore, $h_{\rm TSSTC}$ denotes the lightest CP-even Higgs boson in the present model and $h_{\rm SM}$ denotes the Higgs boson in the SM. In the present model, the yukawa coupling between the top quark and the Higgs boson becomes larger than the SM one since $h_{\rm TSS}$ is composed of the top quark and its vector-like partner. Thus it is possible to obtain the enhancement of the gluon fusion production process: $\sigma_{\rm ggF}(h_{\rm TSSTC})/\sigma_{\rm ggF}(h_{\rm SM})\simeq 2$ in Eq.(\ref{signalstrength-ggF}). However, we find that the $hVV$-coupling in the present model for Eq.(\ref{MTSS0-sample3}) is much smaller than the SM Higgs boson case. This is so, since $v_2/v_{\rm EW} = 0.1$ is satisfied for the case Eq.(\ref{MTSS0-sample3}). This fact ensures that we need not take into account the production process via the vector boson fusion in Eq.(\ref{signalstrength-ggF}) when we compare $\mu_X$ $(X=\gamma\gamma,WW^*,ZZ^*)$ to the LHC data. However, the signal strength of the fermionic decay mode is given by
\beq
\mu_{bb} 
\equiv 
\frac{\sigma_{\rm WH}(h_{\rm TSSTC}) + \sigma_{\rm ZH}(h_{\rm TSSTC})}{\sigma_{\rm WH}(h_{\rm SM}) + \sigma_{\rm ZH}(h_{\rm SM})}
\times
\frac{\text{Br} (h_{\rm TSSTC} \to b\bar{b})}{\text{Br} (h^{\rm SM} \to b\bar{b})}
\,,
\eeq
and we find that the suppression factor $v_2/v_{\rm EW} = 0.1$ affects the production cross section part in the signal strength. Furthermore, we should bear in mind that the suppression factor $v_2/v_{\rm EW} = 0.1$ also affects the decay process which includes the $hVV$-coupling. %
All signal strengths in the present model are shown in the left panel in Fig.\ref{higgs-LHC} as a function of $\epsilon_b$. For comparison, we also present  the values of 
$\mu_{\gamma \gamma , ZZ^* , WW^*}$ from the ATLAS and  CMS data \cite{:2012gk,:2012gu,CMS-PAS-HIG-12-015,CMS-PAS-HIG-12-016}. One can see that all signal strengths are very small ($\mu_X \ll 1$) in the present model due to the above suppression factor if $\epsilon_b \simeq 0.1 - 0.6$. However, $\mu_{VV^*}$ changes drastically for $\epsilon_b \simeq 0.8$. This is because a large $\epsilon_b$ around $\epsilon_b =1$implies that the bottom-yukawa coupling in the present model is smaller than the SM one and its smallness brings i) small $\text{Br} (h_{\rm TSSTC} \to b\bar{b})$ and ii) large $\text{Br} (h_{\rm TSSTC} \to VV^*)$ compared to the SM Higgs boson case.
%

In the present model it is possible to explain the observed diphoton excess, without influencing the other decay channels, by adding a color-singlet isotriplet vector meson $\rho^{0,\pm}_\mu$. In fact, if this vector meson couples to $h_{\rm TSS}$ as ${\cal L}_{h\rho\rho} = g_{h\rho \rho} \cdot h \rho^{+ \mu} \rho^-_\mu$, then the decay width $\Gamma(h_{\rm TSSTC} \to \gamma \gamma)$ gets a contribution
\beq
&&
\Gamma(h_{\rm TSSTC} \to \gamma \gamma) 
\simeq 
\frac{\alpha^2 g^2}{1024 \pi^3}\frac{m^3_h}{M^2_W}
(-7)^2 \times 
\left( \cos \phi \sin (\beta - \alpha) + C_{h\rho\rho} \right)^2
\label{diphoton-decaywidth}
\,,
\eeq
where $C_{h \rho \rho} \equiv [g_{h\rho \rho}/(gM_W)] \times M^2_W/M^2_\rho$. Here we impose that the isotriplet vector meson mass is much larger than $m_h$: $M_\rho \gg 2 m_h$ and $\tan \alpha \simeq 1$, which is the mixing angle in the CP-even higgs boson sector in the present model. In Eq.(\ref{diphoton-decaywidth}) we present the approximate expression for the vector boson contributions to the diphoton decay channel and drop the fermion contributions  since the fermion contributions are smaller than the vector boson contributions. Now, we do not specify the origin of the isotriplet vector meson and we treat $(C_{h \rho \rho}, M_\rho)$ as free parameter. In the right panel in Fig.\ref{higgs-LHC}, we show the signal strength $\mu_X$ for $C_{h\rho\rho} = 0.4$ and $M_\rho = 1 \TeV$ as a function of $\epsilon_b$. We find that this modification, i.e. adding ${\cal L}_{h\rho\rho}$, gives a large contribution to $\Gamma(h_{\rm TSSTC} \to \gamma \gamma)$ but $\text{Br}(h_{\rm TSSTC} \to WW^*/ZZ^*/b\bar{b})$ are not affected by this addition as expected. Therefore it is possible to explain the large diphoton excess in the present model with keeping other decay channels be consistent with the LHC SM Higgs data.

%
\section{Summary}
%
In this talk we have considered a model which involves both the top-seesaw and the technicolor simultaneously. Especially, we have focused on the top-seesaw sector which provides the SM Higgs-like scalar boson by the NJL dynamics. We conclude that it may be possible to realize the $126 \GeV$ Higgs boson in such  model by sharing the top quark mass with another dynamical sector. Furthermore we have found that the Higgs boson in the top-seesaw assisted technicolor model may be consistent with the LHC data if the new isotriplet vector meson exists and couples to the Higgs boson in the top-seesaw assisted technicolor model.


\end{document}